\documentclass{article}
\usepackage{spconf,amsmath,graphicx}
\usepackage{multirow}
\usepackage{float} 
\usepackage{titlesec}
\usepackage{balance}

    \title{Reverse Attention for Lightweight Speech Enhancement on Edge Devices}

\name{Shuubham Ojha$^{1}$, Felix Gervits$^{2}$, Carol Espy-Wilson$^{1}$}
\address{
    $^{1}$ Dept. of Electrical and Computer Engineering \\
    University of Maryland,  College Park, MD, USA 
    $^{2}$ DEVCOM,
    Army Research Lab, \\
    Boston, MA, USA
}

\titleformat{\section}
  {\normalfont\Large\bfseries\centering} 
  {\thesection}                           
  {1em}                                   
  {}                                      

\titlespacing*{\section}{0pt}{1ex}{0.5ex}
\titlespacing*{\subsection}{0pt}{0.8ex}{0.4ex}

\begin{document}

\maketitle

\begin{abstract}
\textbf{This paper introduces a lightweight deep learning model for real-time speech enhancement, designed to operate efficiently on resource-constrained devices. The proposed model leverages a compact architecture that facilitates rapid inference without compromising performance. Key contributions include infusing soft attention-based attention gates in the U-Net architecture which is known to perform well for segmentation tasks and is optimized for GPUs. Experimental evaluations demonstrate that the model achieves competitive speech quality and intelligibility metrics, such as PESQ and Word Error Rates (WER), improving the performance of similarly sized baseline models. We are able to achieve a 6.24\% WER improvement and a 0.64 PESQ score improvement over un-enhanced waveforms.}  


\end{abstract}


\section{Introduction}
\label{sec:intro}
\begingroup
\renewcommand{\thefootnote}{} 
\footnotetext{This material is based on research that is in part supported by the Army
Research Laboratory, Grant No. W911NF2120076.}
\endgroup
Real-time speech enhancement is a critical component in numerous applications, including mobile communication, hearing aids, and voice-controlled systems. The primary objective is to improve the intelligibility and quality of speech signals by reducing background noise and reverberation, thus facilitating clearer communication in challenging acoustic environments. Achieving this in real time necessitates the development of lightweight models capable of processing audio streams with minimal latency and computational overhead. Recent advancements have led to the emergence of compact deep learning architectures tailored for real-time speech enhancement. One such system is Deep Filter Net 3 \cite{9747055} with 2.3 M parameters \cite{s25030630} which is designed to have low processing times for use in hearing aids. 

The significance of real-time speech enhancement extends beyond mere noise reduction. In mobile communication, background noise can severely degrade speech intelligibility, leading to misunderstandings and communication breakdown. In hearing aids, real-time processing is essential to adapt to dynamic acoustic environments, ensuring users receive clear and intelligible speech. Furthermore, in applications such as voice-controlled systems and automatic speech recognition, real-time enhancement is crucial to maintain system responsiveness and accuracy.




This paper explores the use of the Attention Gating (AG) mechanism to extract coarse scale information for use in gating to disambiguate irrelevant and noisy responses in skip connections. We also explore the use of reverse attention (RA) networks for speech enhancement. RA is an advanced mechanism that encourages neural networks to focus on the less confident or misclassified regions of a feature map by explicitly attending to the complement of the current attention response. Unlike traditional attention, which enhances the most salient regions of an image, RA inverts the focus - it suppresses high-response areas and guides the model to concentrate on background regions, weak object boundaries, and previously ignored details. This is especially valuable in segmentation tasks, where the precise delineation of object boundaries and the identification of hard-to-detect regions are critical for performance. 



Our key contribution in this paper is to explore RA and AG approaches in conjunction with each other to improve the ability of the enhancement model to capture fine-grained details that are often lost in traditional architectures. We tune the number of channels in our proposed architecture so that the size of our model is comparable to other well-known models that are designed for edge devices. Finally, we train our model and the baseline models on an in-house dataset that is a mix of everyday and military noises.

\section{Background}
\label{sec:format}

Existing approaches to speech enhancement can be broadly divided into predictive and generative. Predictive methods are generally based on discriminative models for image segmentation. Noisy spectrograms are viewed as images upon which a segmentation task has to be performed to classify speech as the "foreground" and noise as the "background", the latter being finally removed to give enhanced speech. This approach uses an encoder-decoder mechanism with GRUs \cite{9619422}, LSTMs \cite{9650978} and more recently attention layers \cite{deng20_interspeech} which help models in focusing on specific areas of interest but at a significant computation and memory cost. These models tend to leave noise artifacts, especially in the presence of background speech at negative signal-to-noise ratios (SNRs), or cause speech deletion by being too aggressive \cite{9747055}.     


Most two-stage generative methods in the literature \cite{ICLR2025_acde98fb}\cite{unknown}\cite{LEI2026111050} cascade a generator to the predictor to regenerate speech deleted by an aggressive predictor. An important limitation of this approach is that these models are large, with several million parameters and not suitable for real-time speech enhancement over edge devices. Consequently, in this work we scale down generative models in model complexity and investigate mechanisms to improve speech enhancement and recognition performance of the smaller vanilla model.

\section{Methodology}
\label{sec:pagestyle}

\begin{figure}
\begin{minipage}[b]{1.0\linewidth}
  \centering
  \centerline{\includegraphics[width=9.5cm]{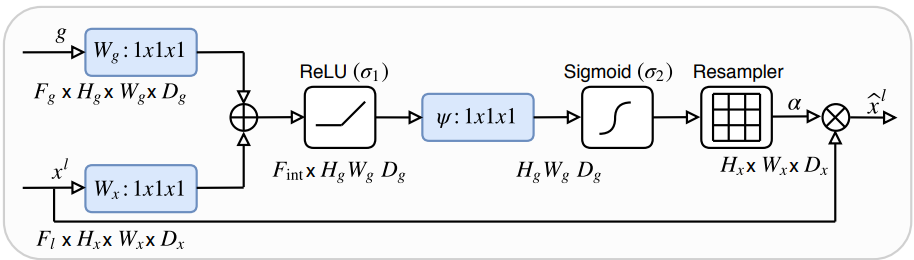}}
  \caption{The Attention Gate from \cite{DBLP:journals/corr/abs-1804-03999}}
\end{minipage}
\end{figure}

\begin{figure*}[t!]
  \centering
  \includegraphics[width=\textwidth]{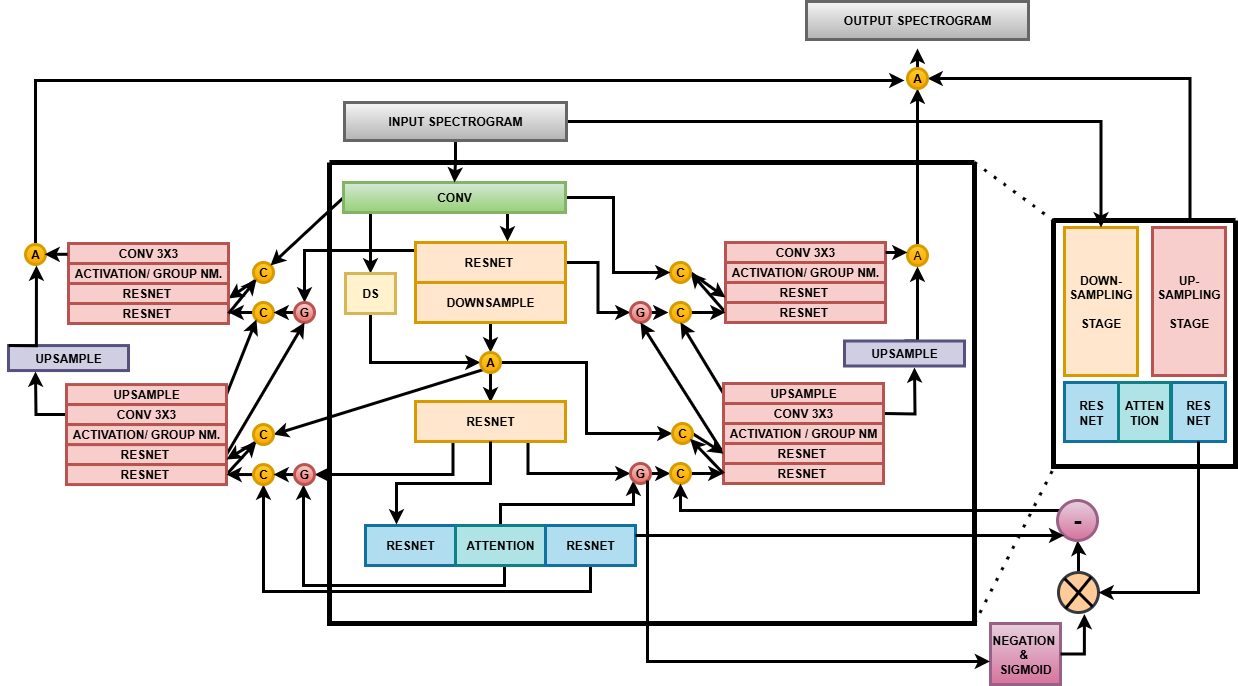}
  \caption{The proposed architecture showing the down sampling stage in beige, up sampling stage in light greyish red and intermediate stage in light blue-green. Reverse attention increases the model width as shown by the box on the extreme right. The negation \& sigmoid, followed by element wise multiplication simulates the modulation in RAN. The network on the right upsamples the negation of input features while that on the left upsamples original features, had there been no RA mechanism.} 
\end{figure*}

\subsection{Attention Gates}
\label{ssec:subhead}
AG is a neural network that allows a model to focus on the most relevant features while suppressing irrelevant ones via adaptive weighing. Recently, AGs have achieved relevance in the field of medical image segmentation. In \cite{DBLP:journals/corr/abs-1804-03999}, the authors were able to achieve significant improvements in locating the pancreas within CT scans. The CT pancreas segmentation problem is particularly challenging owing to the low tissue contrast of the pancreas in CT scans and high variability in organ size among different subjects. Their gate design is presented in Fig. 1. $x^{l}$ represents the input feature map (in our case this comes from the decoder end of the U-Net). $g$ represents a gating signal (in our case from the encoder end), which signifies the focus regions for subsequent decoding. It contains contextual information to emphasize the important locations in the input features.  As shown in the figure, the input features and gating signal are input to the attention gate to obtain attention coefficients $\alpha$. The final output is the element-wise product of $\alpha$ and $x^{l}$. 

The robustness of AGs led to its use in the domain of speech enhancement \cite{deng20_interspeech}, \cite{e25040628}. In this work, we employ AGs over skip connections between the encoder and decoder ends of the U-Net backbone and demonstrate performance improvements compared to the vanilla U-Net architecture.


\subsection{Reverse Attention}

RA enhances a model's ability to capture fine-grained details, especially around object boundaries and ambiguous regions. Traditional segmentation models often struggle in areas where object features are weak or blended with the background, leading to coarse or imprecise results. RA addresses this by explicitly teaching the model to learn not only what the object is, but also what it is not. The idea, introduced in the Reverse Attention Network (RAN), involves a three-branch architecture: a direct prediction branch that learns standard object features, a reverse branch that is trained on the inverse of the object (essentially background or non-object regions), and a RA branch that focuses on confusing or low-confidence areas identified by the model. This third branch amplifies weak activations in those regions, guiding the network to refine its segmentation output where it initially struggled. This is shown in Fig 3. 

\begin{figure}[H]
\begin{minipage}[b]{1.0\linewidth}
  \centering
  \centerline{\includegraphics[width=9.5cm]{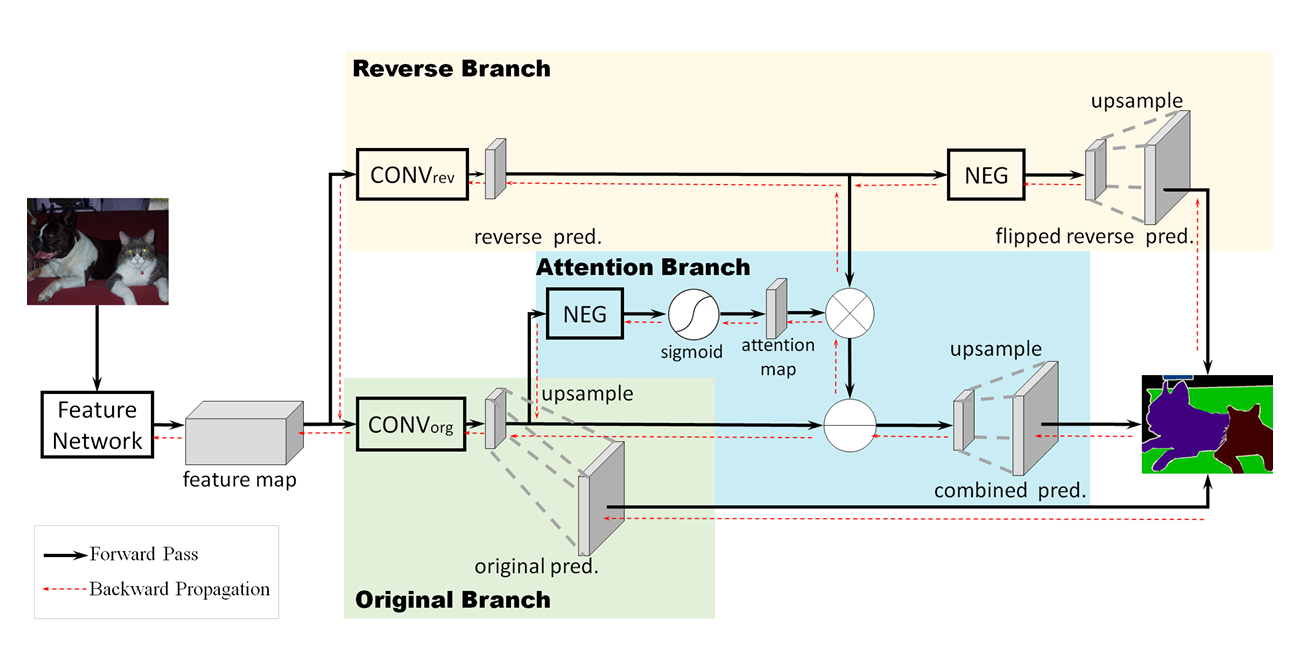}}
  \caption{The Reverse Attention Network from \cite{BMVC2017_18}}\medskip
\end{minipage}
\end{figure}

\subsection{Proposed Architecture}

The proposed architecture is presented in Fig 2. The basic architecture is given by the central black outline. The encoder on the left is a series of four cascaded Resnet layers, each followed by a downsampling stage, which downsamples by a factor of two. The intermediate stage is given two Resnet layers with an Attention layer in between to derive context (the blue - green boxes). The attention gates are shown by circled $\textbf{G}$ whereas the circled $\textbf{C}$ is a combiner that combines the gate output with the output of the corresponding upsampling stage. We realize the combiner as an adder. In our implementation, the gating signal comes from a deeper stage of the U-Net decoder while the input feature comes from a shallower stage of the encoder. The circled \textbf{A} in yellow denote an adder which adds the output of the 3 X 3 convolution layers in adjacent upsampling stages. 

The input and output spectrogram are two channel tensors each, which denote the real and imaginary parts of the complex STFT. The first convolution layer, shown in light green, increases the number of channels to a tunable number via 1 X 1 convolution to obtain detailed abstract features. The wider tensor propagates through the network until it is collapsed into a two channel tensor again by the 3 X 3 convolution layer on the up sampling side. The outputs of the 3 x 3 convolution is upsampled via (purple) upsamplers and added together to produce a two channel output spectrogram. Upsampling by a factor of two is performed to compensate for the prior down sampling, since only tensors of the same size can be added.     

In our experiments we set the light green convolution layer to increase the number of channels from 2 to 16 and 32 respectively. Further we use a model with depth 4, meaning that we use 4 X (Resnet + Downsamplers) stacked together on the left hand side and 4 X (Resnet + Activation + Convolution layers) stacked on the right hand side. However, due to space limitations we present an abridged illustration where we show only 2 stages on each side. \par

The black outline box on the right is a repetition of the central box. Its encoder output is multiplied with the negation of the central encoder passed through the sigmoid activation function. This simulates the modulation of the (negated) reverse features by the original encoded feature as in Fig 3. The residual between positive and negative features is fed to the middle decoder, which corresponds to the central branch in the Fig 3. The decoder on the left is the original decoding branch had there been no reverse mechanism. This corresponds to the lower branch in Fig 3. The right-most decoding branch upsamples the negation of the original features which simulates the uppermost branch of Fig 3.  The output of the three decoders is added to generate a segmented estimate of the clean-speech spectrogram. 


\section{Experiments}
\label{sec:typestyle}

\subsection{Dataset Preparation}

To train the speech enhancement models, 19,992 audio waveforms were taken from the train split of Librispeech \textit{train-clean-100}. These files were divided equally among 27 noises, which ranged from everyday noises such as baby cries to ambient noises from military settings such as helicopters. For each noise type, an equal number of waveforms were randomly chosen to be mixed with it at 5 different SNRs: -3dB, 0dB, 3dB, 6dB and 12dB. These SNRs were selected to model the wide range of scenarios that a general-purpose ASR system might encounter \cite{10974078}. Owing to the relatively large inference time for diffusion models such as StoRM, we kept our validation set to a minimum. The validation set consisted of 10 waveforms from the Librispeech validation set. The set was divided into 5 groups of 2, each for one SNR. The noises used for each set were from the office and crowd domains, as background chatter is considered to be one of the most challenging in the domain of speech enhancement due to the irregular structure of these noises. For the test set, we used 10 noises from the training set. Each noise was combined with 20 utterances from the Librispeech test set at the same five SNRs. Thus, the total number of utterances in the test set was 1000, totaling around 4 hours in duration.

\subsection{Performance Evaluation}    

We trained models for at most 200 epochs or convergence with a patience of 50, whichever is earlier. We evaluate the performance of our model against the NCSNpp-based denoiser only model from \cite{10180108}, trained in the 32 channel configuration that makes it comparable to our model in parameter count. We use our model (from Fig 2) as a backbone for the two-stage configuration, with predictor and generator cascaded together as done by the authors in \cite{10180108}.  We find that the NCSNpp-based predictor-only architecture with 32 channels (parameter count 1.7M) is comparable to our generator + predictor model with 16 channels (parameter count 2.3M). We used both models in depth 4 configuration, as our experiments indicated that increasing depth in general causes the \% WER to increase and PESQ scores to decline.    

\vspace{-0.25cm}
\begin{table}[h!]
\caption{Average PESQ (speech quality) scores at different SNRs for proposed architecture (16 channel configuration), NCSNpp predictor (32 channel configuration) and trained DeepFilterNet 3 models.}
\centering
\begin{tabular}{|c|c|c|c|c|c|}
\hline
\multirow{2}{*}{\textbf{SNR}} & \multicolumn{2}{c|}{\textbf{Ours}} & \multirow{2}{*}{\textbf{Input}} & \multirow{2}{*}{\textbf{NCSNpp}} & \multirow{2}{*}{\textbf{DFNet3}} \\ \cline{2-3}
                       & \textbf{w. RA} & \textbf{w AG} &                         &                         &                         \\ \hline
-3 & 1.7684 & 1.6942 & 1.2661 & 1.3889 & \textbf{1.9055} \\ \hline
0 & 2.0322 & 1.9221 & 1.3174 & 1.4851 & \textbf{2.1701} \\ \hline
3 & 2.2935 & 2.1549 & 1.3948 & 1.6233 &  \textbf{2.4562}\\ \hline
6 & 2.5435 & 2.3972 & 1.4705 & 1.7766 & \textbf{2.7388} \\ \hline
12 & 2.9996 & 2.8324 & 1.59 & 2.194 & \textbf{3.3037} \\ \hline
\end{tabular}

\end{table}

To demonstrate the efficacy of reverse attention we also calculate the \% Word Error Rate (WER) and Perceptual Evaluation of Speech Quality (PESQ) scores without the reverse attention mechanism. We observe that the use of attention gates only provides an improvement of about 4\% in the WER metric compared to the NCSNpp architecture, and the use of reverse attention gives an additional 3\%, averaged across noise types and SNRs. Further, we compare our metrics with Deep Filter Net 3, a well known, light weight 2.13 M parameter model with low processing times. We train and test DF Net 3 on the same dataset as described in section 4.1. Our model outperforms DF Net 3 on the \% WER metric at all SNRs. We attribute this to the aggressive noise suppression tendency of DF Net 3 which causes speech deletion, leading to missing phonemes (Fig. 4). 

\vspace{-0.5cm}

\begin{table}[h!]
\caption{Average \% WER (speech quality) scores at different SNRs for proposed architecture (16 channel configuration), NCSNpp predictor (32 channel configuration) and trained DeepFilterNet 3 models.}
\centering
\begin{tabular}{|c|c|c|c|c|}
\hline
\multirow{2}{*}{\textbf{SNR}} & \multicolumn{2}{c|}{\textbf{Ours}} & \multirow{2}{*}{\textbf{NCSNpp}} & \multirow{2}{*}{\textbf{DeepFilterNet 3}} \\ \cline{2-3}
                       & \textbf{w. RA }  & \textbf{w AG} &                         &                         \\ \hline
-3 & \textbf{24.19} & 24.43 &  37.09 & 41.5 \\ \hline
0 & \textbf{13.58} & 14.59 & 22.06 & 27.9 \\ \hline
3 & \textbf{6.87} & 10.43 & 12.65 & 18.97 \\ \hline
6 & \textbf{2.76} & 8.04 & 8.5 & 13.87 \\ \hline
12 & \textbf{2.11} & 7.43 & 5.29 & 12.66 \\ \hline
\end{tabular}
\end{table}


\vspace{0.35cm}
\section{Discussion}
In this work, we focused on enhancement for resource-constrained hardware such as smartphones or hearing aids, which require low memory, processing time and parameter count ($<$ 10 M) to perform with limited energy consumption \cite{Pal2024SpeechED}\cite{dementyev2025submillisecondlatencyrealtimespeech}. An important limitation of our design is that even though the use of RA improved the \% WER and PESQ scores compared to the AG only configuration, the use of RA almost doubles the number of trainable parameters of model. It also has a longer processing time, especially when used in the predictor + generator mode. However, even with this increased parameter count, our model's parameter count is comparable to Deep Filter Net 3, a well-known light weight speech enhancement model designed for use in hearing aids. DF Net 3's use of a specialized filter bank (ERB filters) in it's preprocessing stage, reduces the frequency resolution of the spectrogram, making it amenable for human ear perception, leading to higher PESQ scores. As a future research direction we would like to integrate DF Net 3's filters with our proposed approach to boost our PESQ scores.

\begin{figure}[ht]
    \centering
    \begin{minipage}{0.45\textwidth}
        \centering
        \includegraphics[width=\linewidth, height=1.5cm]{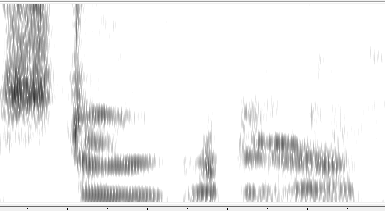}
        \label{fig:fig1}
    \end{minipage}%
    \hfill
    \begin{minipage}{0.45\textwidth}
        \centering
        \includegraphics[width=\linewidth, height=1.2cm]{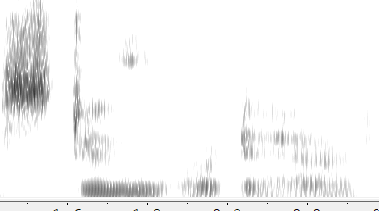}
        \caption{Output from our proposed generative model, PESQ score:1.64 (above) and DF Net 3, PESQ score:1.98 (below). Speech weakening is apparent. PESQ is for waveforms longer than the segments above.}
        \label{fig:fig2}
    \end{minipage}
\end{figure}

\vspace{-0.3cm}

\section{Conclusions}

In this paper, we present a light weight speech enhancement architecture, employing attention gates over skip connections and integrating the reverse attention network on the U-Net backbone. We produced improvements in both ASR loss and perceptibility  (PESQ scores) of the enhanced waveforms with respect to to the predictor-only model output as well as unenhanced waveforms. We also outperform DF Net 3, a light weight speech enhancement baseline on the \% WER metric.

\newpage
\balance
\bibliographystyle{IEEEbib}
\bibliography{bibliography}

\end{document}